\documentclass{article}
\usepackage{ijcai20}

\usepackage{times}
\usepackage{soul}
\usepackage{url}
\usepackage[hidelinks]{hyperref}
\usepackage[utf8]{inputenc}
\usepackage[small]{caption}
\usepackage{multirow}
\usepackage{caption}
\usepackage{subcaption}
\usepackage{graphicx}
\usepackage{xspace}
\graphicspath{{figures/}}
\usepackage{mathtools,commath,amsmath,amssymb,amsfonts,bm}\interdisplaylinepenalty=2500
\usepackage{amsthm}
\usepackage{algpseudocode,algorithm}
\usepackage{tabularx,adjustbox}
\newcolumntype{L}[1]{>{\hsize=#1\hsize\raggedright\arraybackslash}X}%
\newcolumntype{R}[1]{>{\hsize=#1\hsize\raggedleft\arraybackslash}X}%
\newcolumntype{C}[1]{>{\hsize=#1\hsize\centering\arraybackslash}X}%
\usepackage{todonotes}
\usepackage{breqn}
\clearpage
\extrafloats{100}
\RequirePackage{tcolorbox}
\tcbuselibrary{breakable}
\tcbset{parbox=false, left=1ex, right=1ex}

\newcommand{\ie}{\textit{i.e.},\xspace}
\newcommand{\etal}{\textit{et~al.}\xspace}
\newcommand{\eg}{\textit{e.g.},\xspace}

\algnewcommand\algorithmicinput{\textbf{Input:}}
\algnewcommand\Input{\item[\algorithmicinput]}
\algnewcommand\algorithmicoutput{\textbf{Output:}}
\algnewcommand\Output{\item[\algorithmicoutput]}
\algnewcommand\algorithmictier{\textbf{Role:}}
\algnewcommand\Tier{\item[\algorithmictier]}

\title{Threats to Federated Learning: A Survey}
\author{
Lingjuan Lyu$^{1*}$
\and
Han Yu$^{2*}$\And
Qiang Yang$^{3,4}$
\affiliations
$^1$Department of Computer Science, National University of Singapore, Singapore\\
$^2$School of Computer Science and Engineering, Nanyang Technological University, Singapore\\
$^3$Department of AI, WeBank, Shenzhen, China\\
$^4$Department of Computer Science and Engineering, Hong Kong University of Science and Technology
\emails
*Corresponding authors: lyulj@comp.nus.edu.sg, han.yu@ntu.edu.sg
}
\begin{document}
\maketitle

\begin{abstract}
With the emergence of data silos and popular privacy awareness, the traditional centralized approach of training artificial intelligence (AI) models is facing strong challenges. Federated learning (FL) has recently emerged as a promising solution under this new reality. Existing FL protocol design has been shown to exhibit vulnerabilities which can be exploited by adversaries both within and without the system to compromise data privacy. It is thus of paramount importance to make FL system designers to be aware of the implications of future FL algorithm design on privacy-preservation. Currently, there is no survey on this topic. In this paper, we bridge this important gap in FL literature. By providing a concise introduction to the concept of FL, and a unique taxonomy covering threat models and two major attacks on FL: 1) poisoning attacks and 2) inference attacks, this paper provides an accessible review of this important topic. We highlight the intuitions, key techniques as well as fundamental assumptions adopted by various attacks, and discuss promising future research directions towards more robust privacy preservation in FL.
\end{abstract}

\section{Introduction}
\label{sec:introduction}
As computing devices become increasingly ubiquitous, people generate huge amounts of data through their day to day usage. Collecting such data into centralized storage facilities is costly and time consuming. Another important concern is data privacy and user confidentiality as the usage data usually contain sensitive information~\cite{abadi2016deep}. Sensitive data such as facial images, location-based services, or health informatioon can be used for targeted social advertising and recommendation, posing the immediate or potential privacy risks. Hence, private data should not be directly shared without any privacy consideration. As societies become increasingly aware of privacy preservation, legal restrictions such as the General Data Protection Regulation (GDPR) are emerging which makes data aggregation practices less feasible~\cite{FL2019}. 

Traditional centralized machine learning (ML) cannot support such ubiquitous deployments and applications due to infrastructure shortcomings such as limited communication bandwidth, intermittent network connectivity, and strict delay constraints~\cite{li2018learning}. In this scenario, federated learning (FL) which pushes model training to the devices from which data originate emerged as a promising alternative ML paradigm~\cite{mcmahan2016federated}. FL enables a multitude of participants to construct a joint ML model without exposing their private training data~\cite{mcmahan2016federated,bonawitz2017practical}. It can handle unbalanced and non-independent and identically distributed (non-IID) data which naturally arise in the real world~\cite{mcmahan2016communication}. In recent years, FL has benefited a wide range of applications such as next word prediction~\cite{mcmahan2016communication,mcmahan2018learning},  visual object detection for safety~\cite{FedVision}, etc.

\subsection{Types of Federated Learning}
Based on the distribution of data features and data samples among participants, federated learning can be generally classified as horizontally federated learning (HFL), vertically federated learning (VFL) and federated transfer learning (FTL)~\cite{yang2019federated}. 

\begin{table}[!t]
\caption{Taxonomy for horizontal federated learning (HFL).}
\label{tbl:H2B_H2C}
\centering
\scalebox{1}{
\begin{tabularx}{\linewidth}{|c|X|X|X|}
\hline
HFL & Number of Participants & FL Training Participation & Technical Capability 
\tabularnewline
\hline
H2B  & small  & frequent & high
\tabularnewline
\hline
H2C & large & not frequent & low
\tabularnewline
\hline
\end{tabularx}}
\end{table}

Under HFL, datasets owned by each participant share similar features but concern different users~\cite{kantarcioglu2004privacy}. In this paper, we further classify HFL into HFL to businesses (H2B), and HFL to consumers (H2C). A comparison between H2B and H2C is listed in Table~\ref{tbl:H2B_H2C}. The main difference lies in the number of participants, FL training participation level, and technical capability, which can influence how adversaries attempt to compromise the FL system. Under H2B, there are typically a handful of participants. They can be frequently selected during FL training. The participants tend to possess significant computational power and sophisticated technical capabilities~\cite{FL2019}. Under H2C, there can be thousands or even millions of potential participants. In each round of training, only a subset of them are selected. As their datasets tend to be small, the chance of a participant being selected repeatedly for FL training is low. They generally possess limited computational power and low technical capabilities. An example of H2C is Google's GBoard application~\cite{mcmahan2018learning}.

VFL is applicable to the cases in which participants have large overlaps in the sample space but differ in the feature space, \ie different parties hold different attributes of the same records~\cite{vaidya2002privacy}.
VFL mainly targets business participants. Thus, the characteristics of VFL participants are similar to those of H2B participants.

FTL deals with scenarios in which FL participants have little overlap in both the sample space and the feature space~\cite{FL2019}. Currently, there is no published research studying threats to FTL models.

\begin{figure}[!b]
\centering 
\includegraphics[width=1\columnwidth]{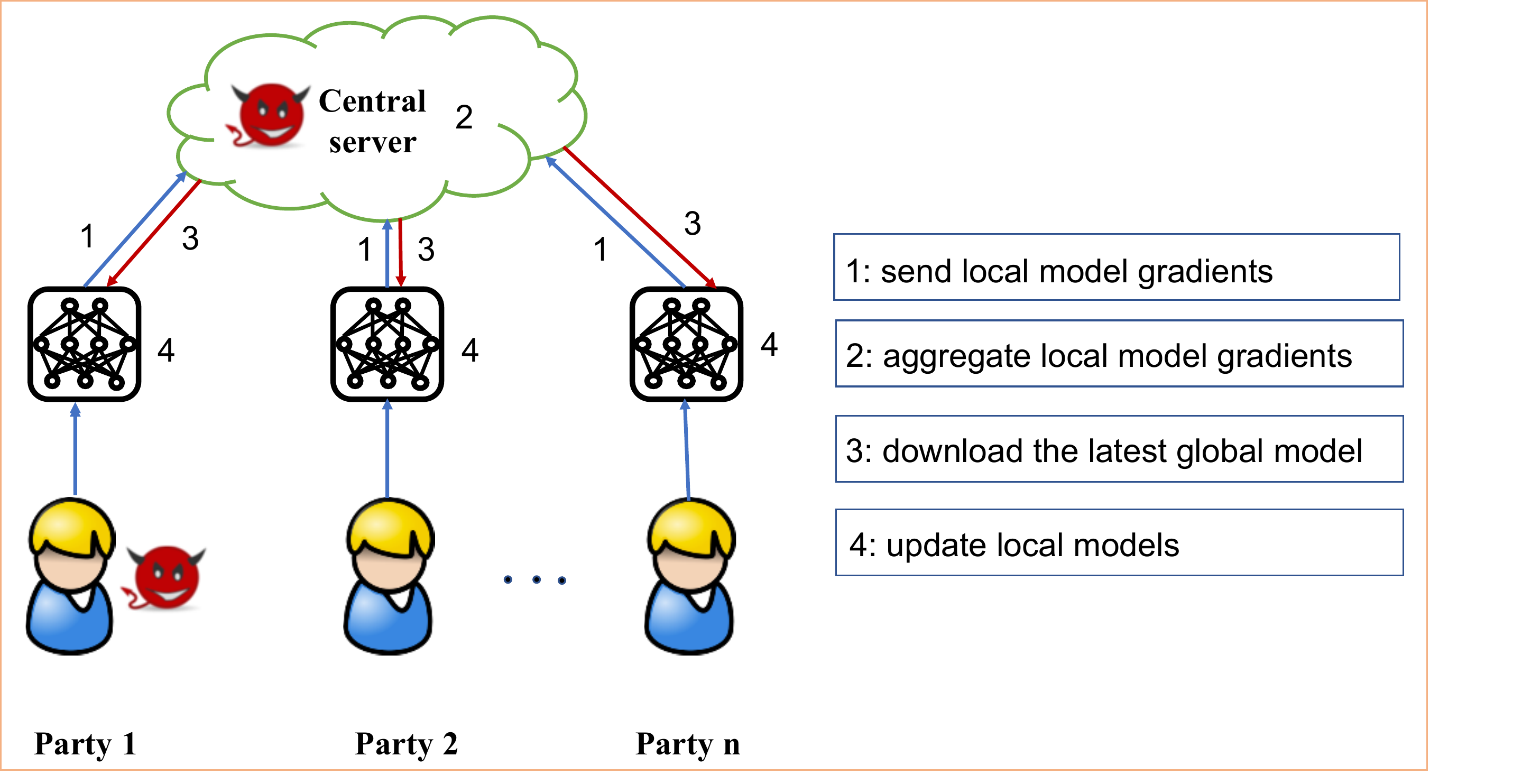}
\caption{A typical FL training process, in which both the (potentially malicious) FL server/aggregator and malicious participants may compromise the FL system.}
\label{fig:FL_train}
\end{figure}

\subsection{Privacy Leakage in FL} 
FL offers a privacy-aware paradigm of model training which does not require data sharing and allows participants to join and leave a federation freely. 
Nevetheless, recent works have demonstrated that FL may not always provide sufficient privacy guarantees, as communicating model updates throughout the training process can nonetheless reveal sensitive information~\cite{bhowmick2018protection,fredrikson2015model,melis2019exploiting} even incur deep leakage~\cite{zhu2019deep}, either to a third-party, or to the central server~\cite{mcmahan2018learning,agarwal2018cpsgd}. For instance, as shown by~\cite{aono2018privacy}, even a small portion of original gradients may reveal information about local data. A more recent work in NeurIPS 2019 showed that the malicious attacker can completely steal the training data from gradients in a few iterations~\cite{zhu2019deep}.

FL protocol designs may contain vulnerabilities for both (1) the (potentially malicious) server, who can observe individual updates over time, tamper with the training process and control the view of the participants on the global parameters; and (2) any participant who can observe the global parameter, and control its parameter uploads. For example, malicious participants can deliberately alter their inputs or introduce stealthy backdoors into the global model. Such attacks pose significant threats to FL. Therefore, it is important to understand the principles behind these attacks. 
Existing survey papers on FL mostly focused on the broad aspect of how to make FL work~\cite{yang2019federated,li2019federated,kairouz2019advances}. In this paper, we survey recent advances in threats to compromise FL to bridge this important gap in the artificial intelligence (AI) research community's understanding in this topic. In particular, we focus on two specific threats initiated by the insiders on FL systems: 1) poisoning attacks that attempt to prevent a model from being learned at all, or to bias the model to produce inferences that are preferable to the adversary; and 2) inference attacks that target participant privacy. The properties of these attacks are summarized in Table \ref{tbl:attacks}.

\begin{table*}[ht]
\caption{A summary of attacks against server-based FL.}
\label{tbl:attacks}
\centering
\resizebox*{1\textwidth}{!}{
\begin{tabular}{|c|c|c|c|c|c|c|c|c|c|}
\hline
\multirow{3}{*}{Attack Type} &\multicolumn{2}{c|}{Attack Targets} & \multicolumn{2}{c|}{Attacker Role}  & \multicolumn{2}{c|}{FL Scenario} & \multicolumn{3}{c|}{Attack Complexity}\\ 
\cline{2-10}
   & Model & Training Data & Participant &Server & H2B & H2C & \multicolumn{2}{c|}{Attack Iteration} & Auxiliary Knowledge Required \\
\cline{8-9}
   & & & & & & &One round &Multiple rounds & \\
\hline
Data Poisoning &YES &NO & YES & NO & YES & YES &YES & YES & YES \\ 
\hline
Model Poisoning &YES & NO & YES & NO & YES & NO &YES & YES & YES \\ 
\hline
Infer Class Representatives &NO &YES  &YES & YES & YES & NO  &NO & YES & YES \\ 
\hline
Infer Membership &NO &YES & YES & YES &YES & NO & NO & YES  & YES \\ 
\hline
Infer Properties &NO & YES & YES & YES & YES & NO &NO &YES &YES  \\ 
\hline
Infer Training Inputs and Labels &NO & YES & NO & YES &YES &NO &NO &YES &NO  \\ 
\hline
\end{tabular}}
\end{table*}

\section{Threat Models}
\label{sec:related_current_work}
Before reviewing attacks on FL, we first present a summary of the threat models. 

\subsection{Insider v.s. Outsider}
Attacks can be carried out by insiders and outsiders. Insider attacks include those launched by the FL server and the participants in the FL system. Outsider attacks include those launched by the eavesdroppers on the communication channel between participants and the FL server, and by users of the final FL model when it is deployed as a service.

Insider attacks are generally stronger than the outsider attacks, as it strictly enhances the capability of the adversary. Due to this stronger behavior, our discussion of attacks against FL will focus primarily on the insider attacks, which can take one of the following three general forms: 
\begin{enumerate}
    \item Single attack: a single, non-colluding malicious participant aims to cause the model to miss-classify a set of chosen inputs with high confidence~\cite{bhagoji2018analyzing,bagdasaryan2018backdoor}; 
    \item Byzantine attack: the byzantine malicious participants may behave completely arbitrarily and tailor their outputs to have similar distribution as the correct model updates, making them difficult to detect~\cite{blanchard2017machine,chen2017distributed,chen2018draco,yin2018byzantine};
    \item Sybil attack: the adversaries can simulate multiple dummy participant accounts or select previously compromised participants to mount more powerful attacks on FL~\cite{fung2018mitigating,bagdasaryan2018backdoor}.
\end{enumerate}

\subsection{Semi-honest v.s. Malicious} 
Under the semi-honest setting, adversaries are considered passive or honest-but-curious. They try to learn the private states of other parties without deviating from the FL protocol. The passive adversaries are assumed to only observe the aggregated or averaged gradient, but not the training data or gradient of other honest participants. Under the malicious setting, an active, or malicious adversary tries to learn the private states of honest participants, and deviates arbitrarily from the FL protocol by modifying, re-playing, or removing messages. This strong adversary model allows the adversary to conduct particularly devastating attacks. 

\subsection{Training Phase v.s. Inference Phase}
Attacks at training phase attempt to learn, influence, or corrupt the FL model itself~\cite{biggio2011support}. 
During training phase, the attacker can run data poisoning attacks to compromise the integrity of training dataset collection, or model poisoning attacks to compromise the integrity of the learning process. The attacker can also launch a range of inference attacks on an individual participant's update or on the aggregate of updates from all participants.

Attacks at inference phase are called evasion/exploratory attacks~\cite{barreno2006can}. They generally do not tamper with the targeted model, but instead, either cause it to produce wrong outputs (targeted/untargeted) or collect evidence about the model characteristics. The effectiveness of such attacks is largely determined by the information that is available to the adversary about the model.
Inference phase attacks can be classified into white-box attacks (i.e. with full access to the FL model) and black-box attacks (i.e. only able to query the FL model). In FL, the model maintained by the server not only suffers from the same evasion attacks as in the general ML setting when the target model is deployed as a service, the model broadcast step in FL renders the model accessible to any malicious client. Thus, FL requires extra efforts to defend against white-box evasion attacks.
In this survey, we omit the discussion of inference phase attacks, and mainly focus on the training phase attacks. 

\section{Poisoning Attacks}
\label{sec:Poisoning}
\begin{figure}[!htp]
\centering 
\includegraphics[width=1\columnwidth]{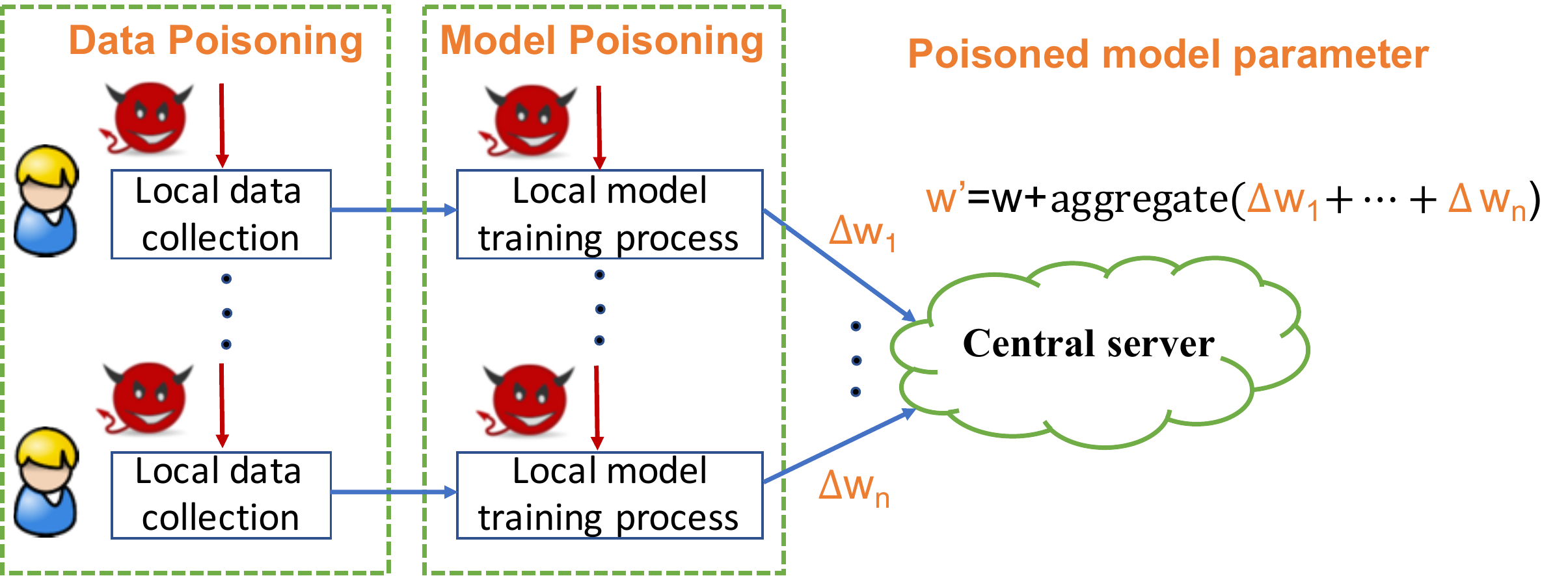}
\caption{Data v.s. model poisoning attacks in FL.}
\label{fig:poisoning_flow1}
\end{figure}

Depending on the attacker's objective, poisoning attacks can be either a) random attacks and b) targeted attacks~\cite{huang2011adversarial}. Random attacks aim to reduce the accuracy of the FL model, whereas targeted attacks aim to induce the FL model to output the target label specified by the adversary. Generally, targeted attacks is more difficult than random attacks as the attacker has a specific goal to achieve.
Poisoning attacks during the training phase can be performed on the data or on the model. Figure~\ref{fig:poisoning_flow1} shows that the poisoned updates can be sourced from two poisoning attacks: (1) data poisoning attack during local data collection; and (2) model poisoning attack during local model training process. At a high level, both poisoning attacks attempt to modify the behavior of the target model in some undesirable way. If adversaries can compromise the FL server, then they can easily perform both targeted and untargeted poisoning attacks on the trained model.

\subsection{Data Poisoning}
Data poisoning attacks largely fall in two categories: 1) clean-label~\cite{shafahi2018poison} and 2) dirty-label~\cite{gu2017badnets}. Clean-label attacks assume that the adversary cannot change the label of any training data as there is a process by which data are certified as belonging to the correct class and the poisoning of data samples has to be imperceptible. In contrast, in dirty-label poisoning, the adversary can introduce a number of data sample it wishes to miss-classify with the desired target label into the training set. 

One common example of dirty-label poisoning attack is the label-flipping attack~\cite{biggio2012poisoning,fung2018mitigating}. The labels of honest training examples of one class are flipped to another class while the features of the data are kept unchanged. For example, the malicious clients in the system can poison their dataset by flipping all 1s into 7s. A successful attack produces a model that is unable to correctly classify 1s and incorrectly predicts them to be 7s. Another weak but realistic attack scenario is backdoor poisoning~\cite{gu2017badnets}. Here, an adversary can modify individual features or small regions of the original training dataset to embed backdoors into the model, so that the model behaves according to the adversary's objective if the input contains the backdoor features (\eg a stamp on an image). However, the performance of the poisoned model on clean inputs is not affected. In this way, the attacks are harder to be detected. 

Data poisoning attacks can be carried out by any FL participant. The impact on the FL model depends on the extent to which participants in the system engage in the attacks, and the amount of training data being poisoned. Data poisoning is less effective in settings with fewer participants like H2C.
 
\subsection{Model Poisoning}
Model poisoning attacks aim to poison local model updates before sending them to the server or insert hidden backdoors into the global model~\cite{bagdasaryan2018backdoor}.

In targeted model poisoning, the adversary’s objective is to cause the FL model to miss-classify a set of chosen inputs with high confidence. Note that these inputs are not modified to induce miss-classification at test time as under adversarial example attacks~\cite{szegedy2013intriguing}. Rather, the miss-classification is a result of adversarial manipulations of the training process. 
Recent works have investigated poisoning attacks on model updates in which a subset of updates sent to the server at any given iteration are poisoned~\cite{blanchard2017machine,bhagoji2018analyzing}. These poisoned updates can be generated by inserting hidden backdoors, and even a single-shot attack may be enough to introduce a backdoor into a model~\cite{bagdasaryan2018backdoor}. 

Bhagoji \etal\shortcite{bhagoji2018analyzing} demonstrated that model poisoning attacks are much more effective than data poisoning in FL settings by analyzing a targeted model poisoning attack, where a single, non-colluding malicious participant aims to cause the model to miss-classify a set of chosen inputs with high confidence. To increase attack stealth and evade detection, they use the alternating minimization strategy to alternately optimize for the training loss and the adversarial objective, and use parameter estimation for the benign participants' updates. This adversarial model poisoning attack can cause targeted poisoning of the FL model undetected. 

In fact, model poisoning subsumes data poisoning in FL settings, as data poisoning attacks eventually change a subset of updates sent to the model at any given iteration~\cite{fung2018mitigating}. This is functionally identical to a centralized poisoning attack in which a subset of the whole training data is poisoned. 
Model poisoning attacks require sophisticated technical capabilities and high computational resources. Such attacks are generally less suitable for H2C scenarios, but more likely to happen in H2B scenarios.

\section{Inference Attacks}
\label{sec:Inference}
\begin{figure}[!t]
\centering
\includegraphics[width=1\columnwidth]{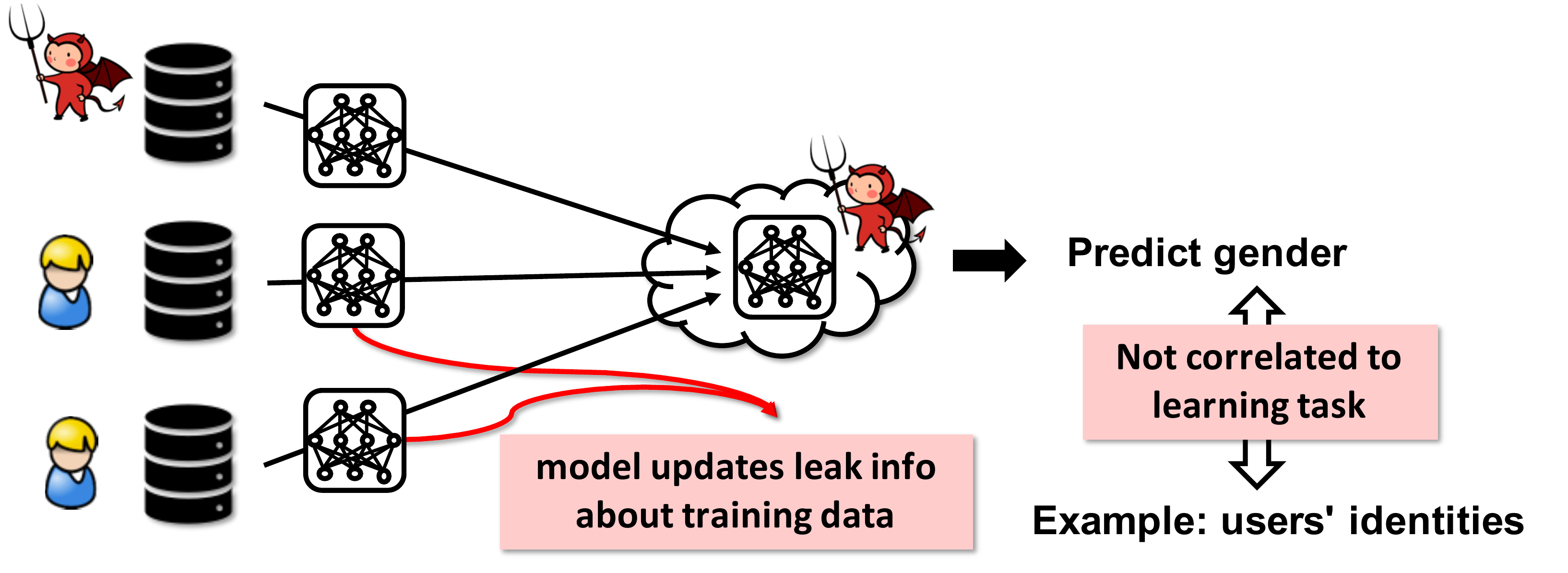}
\caption{Attacker infers information unrelated to the learning task.}
\label{fig:centralized}
\end{figure}

\begin{figure}[!t]
\centering
\includegraphics[width=1\columnwidth]{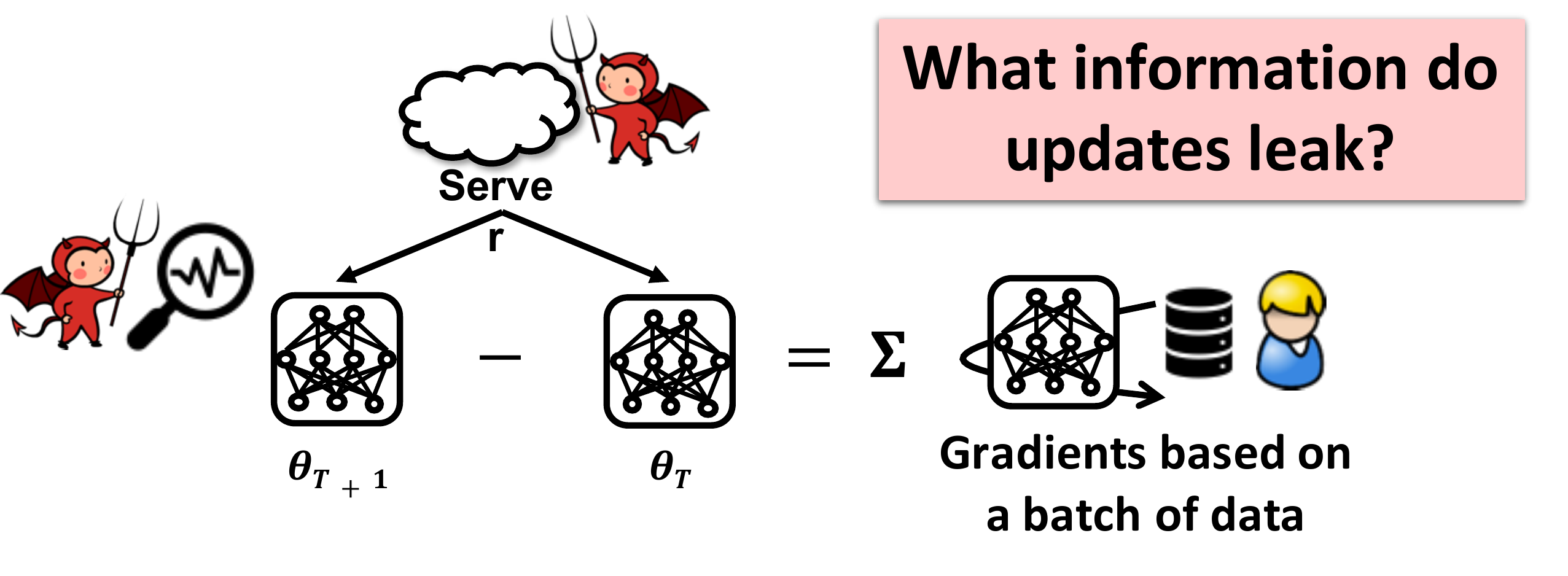}
\caption{Attacker infers gradients from a batch of training data.}
\label{fig:inference_CL}
\end{figure}

Exchanging gradients during FL training can result in serious privacy leakage~\cite{phong2018privacy,su2018securing,melis2019exploiting,zhu2019deep}. As illustrated in Figure~\ref{fig:centralized}, model updates can leak extra information about the unintended features about participants' training data to the adversarial participants, as deep learning models appear to internally recognize many features of the data that are not apparently related with the main tasks.
The adversary can also save the snapshot of the FL model parameters, and conduct property inference by exploiting the difference between the consecutive snapshots, which is equal to the aggregated updates from all participants less the adversary (Figure~\ref{fig:inference_CL}).

The main reason is that the gradients are derived from the participants' private data. In deep learning models, gradients of a given layer are computed using this layer's features and the error from the layer above. In the case of sequential fully connected layers, the gradients of the weights are the inner products of the error from the layer above and the features. Similarly, for a convolutional layer, the gradients of the weights are convolutions of the error from the layer above and the features~\cite{melis2019exploiting}. Consequently, observations of model updates can be used to infer a significant amount of private information, such as class representatives, membership as well as properties associated with a subset of the training data. Even worse, an attacker can infer labels from the shared gradients and recover the original training samples without requiring any prior knowledge about the training set~\cite{zhu2019deep}. 

\subsection{Inferring Class Representatives}
Hitaj \etal~\shortcite{hitaj2017deep} devised an active inference attack called \emph{Generative Adversarial Networks} (GAN) attack on deep FL models. Here, a malicious participant can intentionally compromise any other participant. 
The GAN attack exploits the real-time nature of the FL learning process that allows the adversarial party to train a GAN that generates prototypical samples of the targeted training data which were meant to be private. The generated samples appear to come from the same distribution as the training data. Hence, GAN attack is not targeted at reconstructing actual training inputs, but only class representatives. It should be noted that GAN attack assumes that the entire training corpus for a given class comes from a single participant, and only in the special case where all class members are similar, GAN-constructed representatives are similar to the training data. This resembles model inversion attacks in the general ML settings~\cite{fredrikson2015model}. However, these assumptions may be less practical in FL. Moreover, GAN attack is less suitable for H2C scenarios, as it requires large computation resources.

\subsection{Inferring Membership}
Given an exact data point, membership inference attacks aim to determine if it was used to train the model~\cite{shokri2017membership}. For example, an attacker can infer whether a specific patient profile was used to train a classifier associated with a disease. FL presents interesting new avenues for such attacks. In FL, the adversary's objective is to infer if a particular sample belongs to the private training data of a single party (if target update is of a single party) or of any party (if target update is the aggregate). For example, the non-zero gradients of the embedding layer of a deep learning model trained on natural-language text reveal which words appear in the training batches used by the honest participants during FL model training. This enables an adversary to infer whether a given text appeared in the training dataset~\cite{melis2019exploiting}. 

Attackers in an FL system can conduct both active and passive membership inference attacks~\cite{nasr2019comprehensive,melis2019exploiting}. In the passive case, the attacker simply observes the updated model parameters and performs inference without changing anything in the local or global collaborative training procedure. In the active case, however, the attacker can tamper with the FL model training protocol and perform a more powerful attack against other participants. Specifically, the attacker shares malicious updates and forces the FL model to share more information about the participants' local data the attacker is interested in. This attack, called gradient ascent attack~\cite{nasr2019comprehensive}, exploits the fact that SGD optimization updates model parameters in the opposite direction of the gradient of the loss.

\subsection{Inferring Properties} 
An adversary can launch both passive and active property inference attacks to infer properties of other participants' training data that are independent of the features that characterize the classes of the FL model~\cite{melis2019exploiting}. Property inference attacks assume that the adversary has auxiliary training data correctly labelled with the property he wants to infer. An passive adversary can only observe/eavesdrop the updates and perform inference by training a binary property classifier. An active adversary can use multi-task learning to trick the FL model into learning a better separation for data with and without the property, and thus extract more information. An adversarial participant can even infer when a property appears and disappears in the data during training (e.g., identifying when a person first appears in the photos used to train a gender classifier). The assumption in property inference attacks may prevent its applicability in H2C.

\subsection{Inferring Training Inputs and Labels}
The most recent work called Deep Leakage from Gradient (DLG) proposed an optimization algorithm that can obtain both the training inputs and the labels in just a few iterations~\cite{zhu2019deep}. This attack is much stronger than previous approaches. It can recover pixel-wise accurate original images and token-wise matching original texts. \cite{zhao2020idlg} presented an analytical approach called Improved Deep Leakage from Gradient (iDLG), which can certainly extract labels from the shared gradients by exploiting the relationship between the labels and the signs of corresponding gradients. iDLG is valid for any differentiable model trained with cross-entropy loss over one-hot labels, which is the general case for classification. 

Inference attacks generally assume that the adversaries possess sophisticated technical capabilities and large computational resources. In addition, adversaries must be selected for many rounds of FL training. Thus, it is not suitable for H2C scenarios, but more likely under H2B scenarios. Such attacks also highlight the need for protecting the gradients being shared during FL training, possibly through mechanisms such as homomorphic encryption~\cite{FL2019}. 

\section{Discussions and Promising Directions}
\label{sec:future}
There are still potential vulnerabilities which need to be addressed in order to improve the robustness of FL systems. In this section, we outline research directions which we believe are promising.

\textbf{Curse of Dimensionality:}
Large models, with high dimensional parameter vectors, are particularly susceptible to privacy and security attacks~\cite{chang2019cronus}. Most FL algorithms require overwriting the local model parameters with the global model. This makes them susceptible to poisoning and backdoor attacks, as the adversary can make small but damaging changes in the high-dimensional models without being detected. Thus, sharing model parameters may not be a strong design choice in FL, it opens all the internal state of the model to inference attacks, and maximize the model's malleability by poisoning attacks. 
To address these fundamental shortcomings of FL, it is worthwhile to explore whether sharing model updates is essential. Instead, sharing less sensitive information (e.g., SIGNSGD~\cite{bernstein2018signsgd}) or only sharing model predictions~\cite{chang2019cronus} in a black-box manner may result in more robust privacy protection in FL. 

\textbf{Threats to VFL:}
In VFL, there may only be one party who owns labels for the given learning task. It is unclear if all the participants have equal capability of attacking the FL model, and if threats to HFL can work on VFL. Most of the current threats still focus on HFL. Thus, threats on VFL, which is important to businesses, are worth exploring.

\textbf{FL with Heterogeneous Architectures:} Sharing model updates is typically limited only to homogeneous FL architectures, \ie the same model is shared with all participants. It would be interesting to study how to extend FL to collaboratively train models with heterogeneous architectures~\cite{Gao-et-al:2019,chang2019cronus}, and whether existing attacks and privacy techniques can be adapted to this paradigm. 

\textbf{Decentralized Federated Learning:}
Decentralized FL where no single server is required in the system is currently being studied~\cite{FL2019,lyu2019towards}. This is a potential learning framework for collaboration among businesses which do not trust any third party. In this paradigm, each party could be elected as a server in a round robin manner. It would be interesting to investigate if existing threats on server-based FL still apply in this scenario. Moreover, it may open new attack surfaces. One possible example is that the last party who was elected as the server is more likely to effectively contaminate the whole model if it chooses to insert backdoors. This resembles the fact in server-based FL models which are more vulnerable to backdoors in later rounds of training nearing convergence. Similarly, if decentralized training is conducted in a ``ring all reduce'' manner, then any malicious participant can steal the training data from its neighbors.

\textbf{Weakness of Current Defense:}
FL with secure aggregation are especially susceptible to poisoning attacks as the individual updates cannot be inspected. It is still unclear if adversarial training can be adapted to FL, as adversarial training was developed primarily for IID data, and it is still a challenging problem how it performs in non-IID settings. Moreover, adversarial training typically requires many epochs, which may be impractical in H2C. 
Another possible defense is based on differential privacy (DP).
Record-level DP bounds the success of membership inference, but does not prevent property inference applied to a group of training records~\cite{melis2019exploiting}. Participant-level DP, on the other hand, 
is geared to work with thousands of users for training to converge and achieving an acceptable trade-off between privacy and accuracy~\cite{mcmahan2018learning}. The FL model fails to converge with a small number of participants, making it unsuitable for H2B scenarios. Furthermore, DP may hurt the accuracy of the learned model, which is not appealing to the industry. Further work is needed to investigate if participant-level DP can protect FL systems with few participants.

\textbf{Optimizing Defense Mechanism Deployment:}
When deploying defense mechanisms to check if any adversary is attacking the FL system, the FL server will need to incur extra computational cost. In addition, different defense mechanisms may have different effectiveness against various attacks, and incur different cost. It is important to study how to optimize the timing of deploying defense mechanisms or the anouncement of deterrence measures. Game theoretic research holds promise in addressing this challenge.

Federated learning is still in its infancy and will continue to be an active and important research area for the foreseeable future.
As FL evolves, so will the attack mechanisms. It is of vital importance to provide a broad overview of current attacks on FL so that future FL system designers are aware of the potential vulnerabilities in their designs.
This survey serves as a concise and accessible overview of this topic, and it would greatly help our understanding of the threat landscape in FL. Global collaboration on FL is emerging through a number of workshops in leading AI conferences\footnote{\url{https://www.ntu.edu.sg/home/han.yu/FL.html}}. The ultimate goal of developing a general purpose defense mechanism robust against various attacks without degrading model performance will require interdisciplinary effort from the wider research community.

\section*{Acknowledgments}
This research is supported by the Nanyang Assistant Professorship (NAP); AISG-GC-2019-003; NRF-NRFI05-2019-0002; and NTU-WeBank JRI (NWJ-2019-007).

\bibliographystyle{named}
\bibliography{survey_biblio}

\begin{thebibliography}{}

\bibitem[\protect\citeauthoryear{Abadi \bgroup \em et al.\egroup
  }{2016}]{abadi2016deep}
Mart{\'\i}n Abadi, Andy Chu, Ian Goodfellow, H~Brendan McMahan, Ilya Mironov,
  Kunal Talwar, and Li~Zhang.
\newblock Deep learning with differential privacy.
\newblock In {\em CCS}, pages 308--318, 2016.

\bibitem[\protect\citeauthoryear{Agarwal \bgroup \em et al.\egroup
  }{2018}]{agarwal2018cpsgd}
Naman Agarwal, Ananda~Theertha Suresh, Felix Xinnan~X Yu, Sanjiv Kumar, and
  Brendan McMahan.
\newblock cpsgd: Communication-efficient and differentially-private distributed
  sgd.
\newblock In {\em NeurIPS}, pages 7564--7575, 2018.

\bibitem[\protect\citeauthoryear{Aono \bgroup \em et al.\egroup
  }{2018}]{aono2018privacy}
Yoshinori Aono, Takuya Hayashi, Lihua Wang, Shiho Moriai, et~al.
\newblock Privacy-preserving deep learning via additively homomorphic
  encryption.
\newblock {\em IEEE Transactions on Information Forensics and Security},
  13(5):1333--1345, 2018.

\bibitem[\protect\citeauthoryear{Bagdasaryan \bgroup \em et al.\egroup
  }{2018}]{bagdasaryan2018backdoor}
Eugene Bagdasaryan, Andreas Veit, Yiqing Hua, Deborah Estrin, and Vitaly
  Shmatikov.
\newblock How to backdoor federated learning.
\newblock {\em CoRR, arXiv:1807.00459}, 2018.

\bibitem[\protect\citeauthoryear{Barreno \bgroup \em et al.\egroup
  }{2006}]{barreno2006can}
Marco Barreno, Blaine Nelson, Russell Sears, Anthony~D Joseph, and J~Doug
  Tygar.
\newblock Can machine learning be secure?
\newblock In {\em ICCS}, pages 16--25, 2006.

\bibitem[\protect\citeauthoryear{Bernstein \bgroup \em et al.\egroup
  }{2018}]{bernstein2018signsgd}
Jeremy Bernstein, Jiawei Zhao, Kamyar Azizzadenesheli, and Anima Anandkumar.
\newblock signsgd with majority vote is communication efficient and fault
  tolerant.
\newblock {\em CoRR, arXiv:1810.05291}, 2018.

\bibitem[\protect\citeauthoryear{Bhagoji \bgroup \em et al.\egroup
  }{2018}]{bhagoji2018analyzing}
Arjun~Nitin Bhagoji, Supriyo Chakraborty, Prateek Mittal, and Seraphin Calo.
\newblock Analyzing federated learning through an adversarial lens.
\newblock {\em CoRR, arXiv:1811.12470}, 2018.

\bibitem[\protect\citeauthoryear{Bhowmick \bgroup \em et al.\egroup
  }{2018}]{bhowmick2018protection}
Abhishek Bhowmick, John Duchi, Julien Freudiger, Gaurav Kapoor, and Ryan
  Rogers.
\newblock Protection against reconstruction and its applications in private
  federated learning.
\newblock {\em CoRR, arXiv:1812.00984}, 2018.

\bibitem[\protect\citeauthoryear{Biggio \bgroup \em et al.\egroup
  }{2011}]{biggio2011support}
Battista Biggio, Blaine Nelson, and Pavel Laskov.
\newblock Support vector machines under adversarial label noise.
\newblock In {\em ACML}, pages 97--112, 2011.

\bibitem[\protect\citeauthoryear{Biggio \bgroup \em et al.\egroup
  }{2012}]{biggio2012poisoning}
Battista Biggio, Blaine Nelson, and Pavel Laskov.
\newblock Poisoning attacks against support vector machines.
\newblock {\em CoRR, arXiv:1206.6389}, 2012.

\bibitem[\protect\citeauthoryear{Blanchard \bgroup \em et al.\egroup
  }{2017}]{blanchard2017machine}
Peva Blanchard, Rachid Guerraoui, Julien Stainer, et~al.
\newblock Machine learning with adversaries: Byzantine tolerant gradient
  descent.
\newblock In {\em NeurIPS}, pages 119--129, 2017.

\bibitem[\protect\citeauthoryear{Bonawitz \bgroup \em et al.\egroup
  }{2017}]{bonawitz2017practical}
Keith Bonawitz, Vladimir Ivanov, Ben Kreuter, Antonio Marcedone, H~Brendan
  McMahan, Sarvar Patel, Daniel Ramage, Aaron Segal, and Karn Seth.
\newblock Practical secure aggregation for privacy-preserving machine learning.
\newblock In {\em CCS}, pages 1175--1191, 2017.

\bibitem[\protect\citeauthoryear{Chang \bgroup \em et al.\egroup
  }{2019}]{chang2019cronus}
Hongyan Chang, Virat Shejwalkar, Reza Shokri, and Amir Houmansadr.
\newblock Cronus: Robust and heterogeneous collaborative learning with
  black-box knowledge transfer.
\newblock {\em CoRR, arXiv:1912.11279}, 2019.

\bibitem[\protect\citeauthoryear{Chen \bgroup \em et al.\egroup
  }{2017}]{chen2017distributed}
Yudong Chen, Lili Su, and Jiaming Xu.
\newblock Distributed statistical machine learning in adversarial settings:
  Byzantine gradient descent.
\newblock {\em Proceedings of the ACM on Measurement and Analysis of Computing
  Systems}, 1(2):44, 2017.

\bibitem[\protect\citeauthoryear{Chen \bgroup \em et al.\egroup
  }{2018}]{chen2018draco}
Lingjiao Chen, Hongyi Wang, Zachary Charles, and Dimitris Papailiopoulos.
\newblock Draco: Byzantine-resilient distributed training via redundant
  gradients.
\newblock {\em CoRR, arXiv:1803.09877}, 2018.

\bibitem[\protect\citeauthoryear{Fredrikson \bgroup \em et al.\egroup
  }{2015}]{fredrikson2015model}
Matt Fredrikson, Somesh Jha, and Thomas Ristenpart.
\newblock Model inversion attacks that exploit confidence information and basic
  countermeasures.
\newblock In {\em CCS}, pages 1322--1333, 2015.

\bibitem[\protect\citeauthoryear{Fung \bgroup \em et al.\egroup
  }{2018}]{fung2018mitigating}
Clement Fung, Chris~JM Yoon, and Ivan Beschastnikh.
\newblock Mitigating sybils in federated learning poisoning.
\newblock {\em CoRR, arXiv:1808.04866}, 2018.

\bibitem[\protect\citeauthoryear{Gao \bgroup \em et al.\egroup
  }{2019}]{Gao-et-al:2019}
Dashan Gao, Yang Liu, Anbu Huang, Ce~Ju, Han Yu, and Qiang Yang.
\newblock Privacy-preserving heterogeneous federated transfer learning.
\newblock In {\em IEEE BigData}, 2019.

\bibitem[\protect\citeauthoryear{Gu \bgroup \em et al.\egroup
  }{2017}]{gu2017badnets}
Tianyu Gu, Brendan Dolan-Gavitt, and Siddharth Garg.
\newblock Badnets: Identifying vulnerabilities in the machine learning model
  supply chain.
\newblock {\em CoRR, arXiv:1708.06733}, 2017.

\bibitem[\protect\citeauthoryear{Hitaj \bgroup \em et al.\egroup
  }{2017}]{hitaj2017deep}
Briland Hitaj, Giuseppe Ateniese, and Fernando P{\'e}rez-Cruz.
\newblock Deep models under the gan: information leakage from collaborative
  deep learning.
\newblock In {\em CSS}, pages 603--618, 2017.

\bibitem[\protect\citeauthoryear{Huang \bgroup \em et al.\egroup
  }{2011}]{huang2011adversarial}
Ling Huang, Anthony~D Joseph, Blaine Nelson, Benjamin~IP Rubinstein, and J~Doug
  Tygar.
\newblock Adversarial machine learning.
\newblock In {\em Proceedings of the 4th ACM workshop on Security and
  Artificial Intelligence}, pages 43--58, 2011.

\bibitem[\protect\citeauthoryear{Kairouz \bgroup \em et al.\egroup
  }{2019}]{kairouz2019advances}
Peter Kairouz, H~Brendan McMahan, Brendan Avent, Aur{\'e}lien Bellet, Mehdi
  Bennis, Arjun~Nitin Bhagoji, Keith Bonawitz, Zachary Charles, Graham Cormode,
  Rachel Cummings, et~al.
\newblock Advances and open problems in federated learning.
\newblock {\em CoRR, arXiv:1912.04977}, 2019.

\bibitem[\protect\citeauthoryear{Kantarcioglu and
  Clifton}{2004}]{kantarcioglu2004privacy}
Murat Kantarcioglu and Chris Clifton.
\newblock Privacy-preserving distributed mining of association rules on
  horizontally partitioned data.
\newblock {\em IEEE Transactions on Knowledge \& Data Engineering},
  (9):1026--1037, 2004.

\bibitem[\protect\citeauthoryear{Li \bgroup \em et al.\egroup
  }{2018}]{li2018learning}
He~Li, Kaoru Ota, and Mianxiong Dong.
\newblock Learning iot in edge: Deep learning for the internet of things with
  edge computing.
\newblock {\em IEEE Network}, 32(1):96--101, 2018.

\bibitem[\protect\citeauthoryear{Li \bgroup \em et al.\egroup
  }{2019}]{li2019federated}
Tian Li, Anit~Kumar Sahu, Ameet Talwalkar, and Virginia Smith.
\newblock Federated learning: Challenges, methods, and future directions.
\newblock {\em CoRR, arXiv:1908.07873}, 2019.

\bibitem[\protect\citeauthoryear{Liu \bgroup \em et al.\egroup
  }{2020}]{FedVision}
Yang Liu, Anbu Huang, Yun Luo, He~Huang, Youzhi Liu, Yuanyuan Chen, Lican Feng,
  Tianjian Chen, Han Yu, and Qiang Yang.
\newblock Fedvision: An online visual object detection platform powered by
  federated learning.
\newblock In {\em IAAI}, 2020.

\bibitem[\protect\citeauthoryear{Lyu \bgroup \em et al.\egroup
  }{2019}]{lyu2019towards}
Lingjuan Lyu, Jiangshan Yu, Karthik Nandakumar, Yitong Li, Xingjun Ma, and
  Jiong Jin.
\newblock Towards fair and decentralized privacy-preserving deep learning with
  blockchain.
\newblock {\em CoRR, arXiv:1906.01167}, 2019.

\bibitem[\protect\citeauthoryear{McMahan \bgroup \em et al.\egroup
  }{2016a}]{mcmahan2016communication}
H~Brendan McMahan, Eider Moore, Daniel Ramage, Seth Hampson, et~al.
\newblock Communication-efficient learning of deep networks from decentralized
  data.
\newblock {\em CoRR, arXiv:1602.05629}, 2016.

\bibitem[\protect\citeauthoryear{McMahan \bgroup \em et al.\egroup
  }{2016b}]{mcmahan2016federated}
H~Brendan McMahan, Eider Moore, Daniel Ramage, and Blaise~Aguera y~Arcas.
\newblock Federated learning of deep networks using model averaging.
\newblock {\em CoRR, arXiv:1602.05629}, 2016.

\bibitem[\protect\citeauthoryear{McMahan \bgroup \em et al.\egroup
  }{2018}]{mcmahan2018learning}
H~Brendan McMahan, Daniel Ramage, Kunal Talwar, and Li~Zhang.
\newblock Learning differentially private recurrent language models.
\newblock In {\em ICLR}, 2018.

\bibitem[\protect\citeauthoryear{Melis \bgroup \em et al.\egroup
  }{2019}]{melis2019exploiting}
Luca Melis, Congzheng Song, Emiliano De~Cristofaro, and Vitaly Shmatikov.
\newblock Exploiting unintended feature leakage in collaborative learning.
\newblock In {\em SP}, pages 691--706, 2019.

\bibitem[\protect\citeauthoryear{Nasr \bgroup \em et al.\egroup
  }{2019}]{nasr2019comprehensive}
Milad Nasr, Reza Shokri, and Amir Houmansadr.
\newblock Comprehensive privacy analysis of deep learning: Passive and active
  white-box inference attacks against centralized and federated learning.
\newblock In {\em SP}, pages 739--753, 2019.

\bibitem[\protect\citeauthoryear{Phong \bgroup \em et al.\egroup
  }{2018}]{phong2018privacy}
Le~Trieu Phong, Yoshinori Aono, Takuya Hayashi, Lihua Wang, and Shiho Moriai.
\newblock Privacy-preserving deep learning via additively homomorphic
  encryption.
\newblock {\em IEEE Transactions on Information Forensics and Security},
  13(5):1333--1345, 2018.

\bibitem[\protect\citeauthoryear{Shafahi \bgroup \em et al.\egroup
  }{2018}]{shafahi2018poison}
Ali Shafahi, W~Ronny Huang, Mahyar Najibi, Octavian Suciu, Christoph Studer,
  Tudor Dumitras, and Tom Goldstein.
\newblock Poison frogs! targeted clean-label poisoning attacks on neural
  networks.
\newblock In {\em NeurIPS}, pages 6103--6113, 2018.

\bibitem[\protect\citeauthoryear{Shokri \bgroup \em et al.\egroup
  }{2017}]{shokri2017membership}
Reza Shokri, Marco Stronati, Congzheng Song, and Vitaly Shmatikov.
\newblock Membership inference attacks against machine learning models.
\newblock In {\em SP}, pages 3--18, 2017.

\bibitem[\protect\citeauthoryear{Su and Xu}{2018}]{su2018securing}
Lili Su and Jiaming Xu.
\newblock Securing distributed machine learning in high dimensions.
\newblock {\em CoRR, arXiv:1804.10140}, 2018.

\bibitem[\protect\citeauthoryear{Szegedy \bgroup \em et al.\egroup
  }{2013}]{szegedy2013intriguing}
Christian Szegedy, Wojciech Zaremba, Ilya Sutskever, Joan Bruna, Dumitru Erhan,
  Ian Goodfellow, and Rob Fergus.
\newblock Intriguing properties of neural networks.
\newblock {\em CoRR, arXiv:1312.6199}, 2013.

\bibitem[\protect\citeauthoryear{Vaidya and Clifton}{2002}]{vaidya2002privacy}
Jaideep Vaidya and Chris Clifton.
\newblock Privacy preserving association rule mining in vertically partitioned
  data.
\newblock In {\em KDD}, pages 639--644, 2002.

\bibitem[\protect\citeauthoryear{Yang \bgroup \em et al.\egroup
  }{2019a}]{yang2019federated}
Qiang Yang, Yang Liu, Tianjian Chen, and Yongxin Tong.
\newblock Federated machine learning: Concept and applications.
\newblock {\em ACM Transactions on Intelligent Systems and Technology (TIST)},
  10(2):1--19, 2019.

\bibitem[\protect\citeauthoryear{Yang \bgroup \em et al.\egroup
  }{2019b}]{FL2019}
Qiang Yang, Yang Liu, Yong Cheng, Yan Kang, Tianjian Chen, and Han Yu.
\newblock {\em Federated Learning}.
\newblock Morgan \& Claypool Publishers, 2019.

\bibitem[\protect\citeauthoryear{Yin \bgroup \em et al.\egroup
  }{2018}]{yin2018byzantine}
Dong Yin, Yudong Chen, Kannan Ramchandran, and Peter Bartlett.
\newblock Byzantine-robust distributed learning: Towards optimal statistical
  rates.
\newblock {\em CoRR, arXiv:1803.01498}, 2018.

\bibitem[\protect\citeauthoryear{Zhao \bgroup \em et al.\egroup
  }{2020}]{zhao2020idlg}
Bo~Zhao, Konda~Reddy Mopuri, and Hakan Bilen.
\newblock idlg: Improved deep leakage from gradients.
\newblock {\em CoRR, arXiv:2001.02610}, 2020.

\bibitem[\protect\citeauthoryear{Zhu \bgroup \em et al.\egroup
  }{2019}]{zhu2019deep}
Ligeng Zhu, Zhijian Liu, and Song Han.
\newblock Deep leakage from gradients.
\newblock In {\em NeurIPS}, pages 14747--14756, 2019.

\end{thebibliography}
\end{document}